\documentclass[12pt]{iopart}

\usepackage{graphicx}

\begin{document} 

\title[Dark Matter in dwarf spheroidal galaxies and indirect detection]{Dark Matter in Dwarf Spheroidal Galaxies and Indirect Detection: A Review}

\author{Louis E. Strigari}
\address{Mitchell Institute for Fundamental Physics and Astronomy,  Department of Physics and Astronomy, Texas A\&M University, College Station, TX 77845, USA}
\ead{strigari@tamu.edu}

%\begin{document} 
%\maketitle 

\begin{abstract}
Indirect dark matter searches targeting dwarf spheroidal galaxies (dSphs) have matured rapidly during the past decade. This has been because of the substantial increase in kinematic data sets from the dSphs, the new dSphs that have been discovered, and the operation of the Fermi-LAT and many ground-based gamma-ray experiments. Here we review the analysis methods that have been used to determine the dSph dark matter distributions, in particular the ``J-factors," comparing and contrasting them, and detailing the underlying systematics that still affect the analysis. We discuss prospects for improving measurements of dark matter distributions, and how these interplay with future indirect dark matter searches. 
\end{abstract} 

\section{Introduction} 
\noindent 
\par Indirect dark matter searches, which aim to identify particle dark matter through its annihilation or decay into Standard Model (SM) particles from astronomical targets, are maturing rapidly~\cite{Porter:2011nv,Buckley:2013bha,Conrad:2015bsa,Gaskins:2016cha}. Many astronomical targets, both within our Galaxy and well-beyond it, have now been studied across a wide range of energies. In combination with direct dark matter searches and terrestrial searches, indirect searches play a crucial role in identifying the nature of dark matter and also beyond the SM physics. Indirect searches connect the macrophysics of the dark matter distribution in galaxies to the microphysics governing dark matter interactions with itself and with the SM, so they depend strongly on accurate measurements of the macroscopic dark matter distribution in galaxies. 

\par The dark matter distributions in galaxies are determined by the kinematics of the luminous stars and gas, which are used to extract the underlying gravitational potentials. Because different physics governs the dynamics of galaxies at different mass and luminosity scales, the kinematics of galaxies can differ substantially, and interpreting the data can be complicated. The most dark matter dominated halos to include clusters are those that lie at the extreme ends of the luminosity distribution: clusters of galaxies at the most luminous and massive end, and ultra-faint dwarf galaxies at the least luminous and massive end. At intermediate mass and luminosity, elliptical galaxies and spiral galaxies like the Milky Way have dark to luminous mass ratios $\sim 30-50$~\cite{Behroozi:2012iw}, and can be significantly baryon dominated in their central regions. 

\par Dwarf spheroidal galaxies (dSphs) lie at the extreme low luminosity end of the galaxy distribution~\cite{Mateo:1998wg,McConnachie:2012vd}, and are particularly intriguing objects that connect the macrophysics of dark matter to its microphysics. Most of the known dSphs lie within the viral radius of the Milky Way ($\sim 300$ kpc), and are thus also satellites that are gravitationally bound to the Galaxy. As deduced from their stellar kinematics, the dSphs are the most dark matter-dominated galaxies known~\cite{Battaglia:2013wqa,Walker:2012td}. They contain very old stellar populations, and are almost entirely devoid of gas, in the most extreme cases showing upper limits on their gas mass of $< 1$ M$_\odot$~\cite{Grcevich:2009gt,2014ApJ...795L...5S}. 

\par While many of the known dSphs are satellite galaxies of the Milky Way, the remainder of them lie just beyond the edge of the Milky Way halo or are satellites of M31. Because they are so faint and diffuse, dSphs are difficult to detect, with the census of Milky Way satellites and dSphs continually growing as new and more sensitive surveys are deployed~\cite{Willman:2009dv}. In the past couple of years, most of the new Milky Way satellites have been discovered in the Dark Energy Survery (DES)~\cite{Bechtol:2015cbp,Drlica-Wagner:2015ufc}, and several of these have been targets of kinematic follow-up~\cite{Simon:2015fdw,Walker:2015mla,Simon:2016mkr}. These kinematic studies are important in distinguishing dark matter-dominated dSphs from globular clusters~\cite{Willman:2012uj}, the latter of which are consistent with being devoid of dark matter.  

%They are also important in determining whether the systems are in dynamical equilibrium or are undergoing tidal interactions with the Milky Way.  

\par The formation of dSphs is still not well understood. The best theoretical method to study the formation of dSphs is with numerical simulations of cosmological structure formation, which  are now able to identify galaxies that have properties similar to the dSphs. Interestingly, the properties of dSphs in the simulation of an environment like the Local Group of galaxies are strong functions of the nature of dark matter, 
 if it is warm~\cite{Lovell:2013ola}, or has a large degree of self-interactions~\cite{Vogelsberger:2012ku,Rocha:2012jg}. The key step in ultimately determining whether these simulations can directly constrain dark matter lies in disentangling the effect of baryons versus the effect that the nature of the dark matter has in the formation of the dSphs~\cite{Sawala:2014hqa,Sawala:2014baa}. 

%\par The measured dark matter distributions can directly constrain the properties of particle dark matter, for instance if phase space constraints manifest themselves on macroscopic scales. The measured dark matter distributions may also provide a necessary ingredient to determine the strength of signals from particle annihilation or decay, which is the case for indirect dark matter detection. 

\par Interpreting the measured kinematics of dSphs, and understanding the theory of how they formed, is important because the dSphs are amongst the most important targets for indirect dark matter detection~\cite{Conrad:2015bsa}. The dSphs are the cleanest targets for indirect detection because of their high dark-to-luminous mass ratios, the fact that they are devoid of gas and have very low intrinsic sources of high energy photons, and that many of them are nearby. Further, in a Lambda-Cold Dark Matter (LCDM) formation scenario, there is expected to be a small contribution to the flux from unresolved dark matter substructure within them~\cite{Springel:2008zz}. This implies that the kinematic data from the dSphs are able to place robust constraints on the dark matter annihilation cross section times the relative velocity, $\sigma v$. 

\par In the past several years many indirect dark matter searches have targeted dSphs, most prominently those that have used gamma-rays in the energy range $\sim 100$ MeV to $\sim 100$ TeV~\cite{Abdo:2010ex,GeringerSameth:2011iw,Ackermann:2011wa,Ackermann:2013yva,Geringer-Sameth:2014qqa,Ackermann:2015zua}. This is an important regime to cover for dark matter if it is associated with new physics at the weak scale~\cite{Jungman:1995df}. Further, for dark matter at this mass range, there is a ``thermal relic" scale for the annihilation cross section, $\sigma v \simeq 3 \times 10^{-26}$ cm$^3$ s$^{-1}$, which for thermally-produced dark matter in the early universe produces a dark matter abundance consistent with the measured value~\cite{Steigman:2012nb}. The searches for gamma-rays from dSphs use nearly the entire population of known dSphs, even extending to the faintest dSphs that are just now being uncovered by surveys such as the DES. 

\par Underlying the interpretation of these indirect dark matter searches is a robust determination of the dark matter potentials of dSphs from their stellar kinematics. While determination of these potentials is in many respects simpler than it is for larger galaxies with more complex stellar and gas kinematics, there are systematic issues that must be dealt with for dSphs. For example, the potential of the dSphs may be non-spherical, they may dynamically evolve as they orbit the Milky Way, and their stars may be tidally stripped. In addition, since the dSphs are the lowest velocity dispersion galaxies that have been identified, they are typically just above the threshold that instruments are able to effectively measure the velocities of the member stars and intrinsic dispersions.

\par This article provides a review of methods for determining dark matter distributions of dSphs, and explores the implications for indirect dark matter searches and for constraints on particle dark matter. Section~\ref{sec:theory} begins by reviewing the theoretical modeling of dSphs, discusses the most prominent modeling methods in the literature, and connects this modeling to the measured annihilation cross section. Section~\ref{sec:data} explores the statistical methods that are used to fit the models to the measured stellar kinematics. Section~\ref{sec:results} reviews the recent important results, primarily focusing on the constraints on the annihilation cross section using gamma-ray telescopes. Section~\ref{sec:future} discusses the prospects for improving upon the current results with forthcoming gamma-ray and astronomical data. 

\section{Theory modeling} 
\label{sec:theory} 
\par In this section we review the theory underlying the determination of the dSph dark matter distributions. We start out with a broad discussion of the basic theoretical principles, and then move on to review how these principles are used in the specific modeling methods. We then discuss how the measurements of the dark matter distributions are used to calculate of the flux of SM particles produced. Throughout the discussion, we emphasize the systematics that must be accounted for, and project forward to the sections that follow to highlight the systematics that are incurred in determination of the annihilation cross section.  

\subsection{Basic theoretical principles} 
\par
\begin{figure}[h]
\includegraphics[width=16cm]{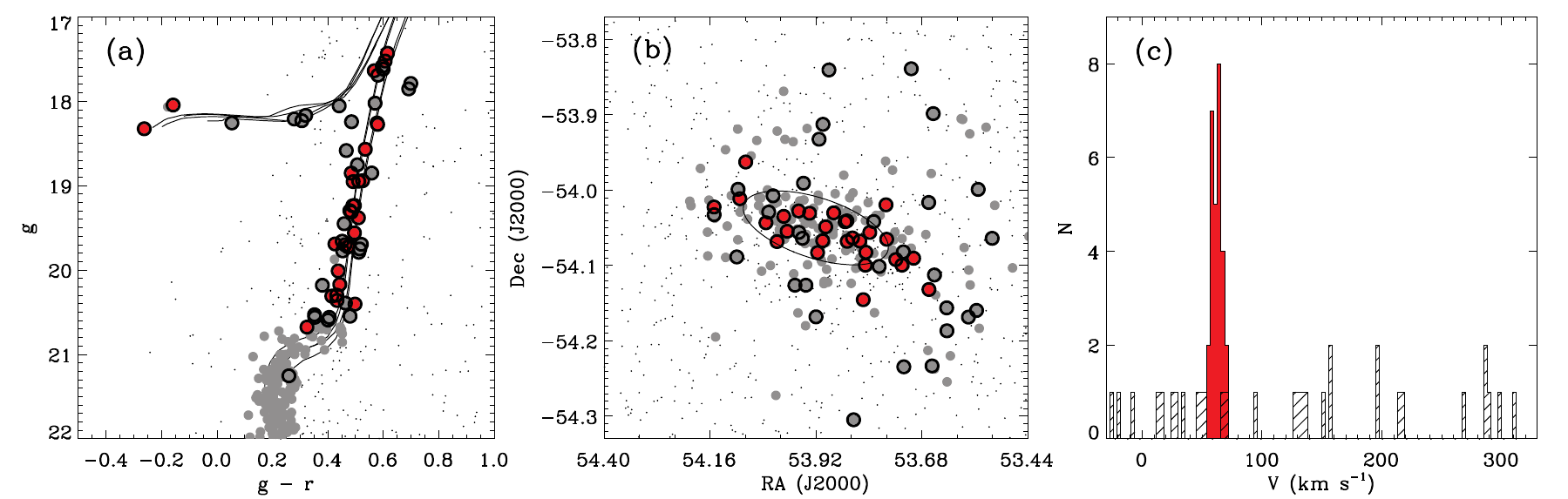}
\caption{Color-magnitude diagram (left), sky position (middle), and velocity distribution (right) of members stars in the Reticulum II dSph. In the left panel, the g-band magnitude is plotted vs. the g-r color. In the left and middle panels, the red dots are identified as member stars of Reticulum II, and the red-filled histogram in the right panel is the velocity distribution of the member stars. The grey points in the left and middle, and the grey hatched regions in the right are foreground and background non-members. The model isochrones in the left panel and the distribution of velocities in the right panel are used to identify the members. In the middle panel, the black ellipse indicates the half-light radius. Figure reproduced from Simon et al.~\cite{Simon:2015fdw}.} 
\label{fig:retII}
\end{figure}

 Stars that are identified as members of a dSph lie along an locus on a color-magnitude diagram (CMD) indicating that they are at nearly the same distance. The shape of the isochrone is determined by parameters that characterize the stellar population, such as age and metallicity. The isochrones that dSph stars lie along are similar to those of globular clusters, indicating that the dSphs contain old stellar populations, with age typically $\sim 10$ Gyrs. This implies that these are among the first generations of stars to have formed when galaxies began to assemble in the early universe. Most of the stars in the dSphs  that are resolved are red giant branch (RGB) stars; these are amongst the most luminous stars that populate a CMD. For dSphs that are sufficiently bright and nearby, main sequence stars can be identified. For a typical dSph, the apparent magnitude of the brightest RGB stars are in the range $m_{\rm v} = 15-18$. Figure~\ref{fig:retII} shows the CMD for Reticulum II, a Milky Way satellite recently discovered by DES. For recent CMDs of the brightest dSphs, see e.g. Refs.~\cite{2014MNRAS.444.3139M,2015MNRAS.453..690B,2016MNRAS.460...30R}.

\par The kinematics of the dSphs are derived from the projected positions and the line-of-sight velocities of resolved member stars. The typical uncertainty on the measured velocity of a star is $\sim 1-2$ km/s, which is precise enough to measure the velocity dispersion in most cases. The two dimensional measure of the projected position and the one-dimensional measure of the velocity represent only three of the six possible phase space coordinates. For a typical dSph distance, the astrometric sensitivity of present instruments is not good enough to measure the tangential component of the velocity of a large sample of stars to the required precision of $\sim$ few km/s (see however the recent work of Ref.~\cite{2017arXiv171108945M}). It is also at present not possible to directly measure the three-dimensional distance of a star relative to the center of the dSph. 

\par Despite these limitations, the fact that the velocities of individual stars in dSphs can be measured makes them particularly unique. Only in the Magellanic Clouds and in M31 is it also possible to obtain kinematic information from the velocities of resolved member stars. For other galaxies, kinematic information is typically obtained through integrated or unresolved starlight.
The measurements of resolved stars in dSphs are similar to those obtained from Milky Way globular clusters and from open star clusters, which have similar velocity dispersions and magnitudes, though they have smaller sizes. This fact implies that the dSphs contain a significant amount of dark matter, whereas globular and open clusters do not require dark matter. 

\par The vast majority of the Milky Way dSphs are devoid of gas; the only exception is Leo T, which is at the extreme edge of Milky Way's dark matter halo, $\sim 420$ kpc from the Galactic center. Due to their lack of gas, a rotation signal from these galaxies must be deduced from the velocities of the resolved stars. For most dSphs, a comparison of the rotational velocity, $v_{rot}$, with the line-of-sight velocity dispersion, $\sigma_{||}$, indicates that $v_{rot}/\sigma_{||} < 1$, implying that they are held in equilibrium by the random motions of the stars~\cite{2017MNRAS.465.2420W}. Amongst the brightest Milky Way dSphs with large velocity samples, only Sculptor shows a possible indication of rotation~\cite{Battaglia:2008jz}. There is, however, clear rotation in AND II, which is one of the brightest satellites of M31. This galaxy has a prolate rotation about its minor axis of  $\sim 8-9$ km/s~\cite{Ho:2012zr}. 

\par The line-of-sight velocity dispersions of the dSphs are measured to be in the range $\sim 3-10$ km/s, and their luminosities vary in the range $L_\star \simeq 10^3 - 10^7$ L$_\odot$. Assuming that the stellar mass-to-light ratios for dSphs are near unity (in units of the solar mass-to-light ratio) which is a good estimation for populations of stars similar to dSphs, the total stellar masses are in the range $M_\star \simeq 10^3 - 10^7$ M$_\odot$. For a stellar system with a measured velocity dispersion of $\sigma_{||}$ and a half-light radius $r_h$, an approximate measurement of the mass at the half-light radius may be obtained from the following estimator~\cite{Walker:2009zp,Wolf:2009tu} 
\begin{equation} 
M \simeq 4 G^{-1} \sigma_{||}^2 r_h.
\label{eq:wolf} 
%M \simeq k G^{-1} \sigma_{los}^2 r_h, 
\end{equation} 
As discussed below, this formula is derived from the Jeans method. It assumes that the system is spherically-symmetric, and the velocity dispersion is nearly constant as a function of radius. For nearly all dSphs it is straightforward to show that the measured $\sigma_{||}$ and $r_h$ imply dark matter mass much greater than the mass deduced in stars. 

\par There are several systematics that must be accounted for in extracting the true velocity dispersion from the measured velocities. First, there are instrumental systematics. High resolution spectrographs have systematic floors which limit the measurement of stellar velocities to greater than $\sim 2$ km/s. This implies that dispersions at or below this floor are difficult to resolve~\cite{Simon:2007dq}. In addition to this instrumental systematic floor, there is a theoretical systematic that arises due to the contribution to the measured velocity dispersion from unseen binary companions. This binary contribution can only be measured directly in cases where stars have multiple velocity measurements at different epochs, and are observed to inflate the velocity dispersion by less than approximately $1$ km/s. For the entire stellar population the binary contribution must be modeled, assuming parameters for the orbital distributions~\cite{Wilkinson:2001ut,Minor:2010vp,Minor:2013bj}. 

\par Despite their name, dSphs are not entirely spherical; the typical projected minor-to-major axis ratio for a dSph is $\sim 0.3$~\cite{1995MNRAS.277.1354I,McConnachie:2005td}. There are no empirical constraints on the shapes of the dark matter halo that the dSph resides in, and it is not yet possible to tell whether the dark matter halo of the dSphs have the same shape and are oriented in the same direction as the stellar distribution. The extent of the dark matter halos are also essentially unconstrained from stellar kinematic data. 

\par In spite of these complications, it is standard practice to model the dark matter distributions as spherically symmetric, for reasons of simplicity and also because spherical models typically provide accurate estimates of the true potential~\cite{Campbell:2016vkb,2017arXiv170605383G}. In addition, in the limit of isotropic orbits in the center of the system, spherical symmetry has been shown to provide a good description of the potential~\cite{Pontzen:2015ova}. Most generally the potential is modeled by the sum of two spherical components, the stars and the dark matter, though for most dSphs the stellar contribution to the potential is negligible. Only near the center of Fornax and possibly Sagittarius do the stars contribute significantly to the potential. The stars and the dark matter are likely not described by the same distributions; for example, the dark matter halo is likely more extended that the stellar distribution. These different spatial distributions of the two components may reflect the formation history of a dSph. 

\par Another theoretical issue that must be considered is the contribution of subhalos within the dSph to the mass profile. This question is again best answered by simulations of Milky Way-like galaxies. The results from dark matter-only simulations show that the shape of the mass function of substructure within subhalos in Milky Way-mass galaxies is similar to the shape of the mass function of subhalos within the Milky Way, though the normalization of the sub-subhalo mass function is reduced by about 50\%~\cite{Springel:2008cc}. This implies that these sub-subhalos contribute a small amount to the dark matter annihilation flux~\cite{Springel:2008zz,Martinez:2009jh}, which will be important for the interpretation of gamma-ray data discussed below.  

\par Hydrodynamic simulations that include baryonic physics now produce satellite galaxies with structural and kinematic properties similar to the dSphs. The simulated satellites that can viably host the dSphs have dark matter distributions that are similar to their counterparts in dark matter-only simulations. However the hydro simulations are most valuable in determining the structure of the stellar distributions, and how these distributions differ from that of the dark matter. Also, they can determine whether there is faint tidal debris associated with a dSph beyond the current photometric thresholds~\cite{Penarrubia:2007zx,2015MNRAS.454.2401B,Wang:2016qol}. Satellites in cosmological simulations with baryons are found to be more spherical than field halos~\cite{Campbell:2016vkb}, which may be explained by tidal stripping~\cite{2015MNRAS.447.1112B}. This is similar to the dark matter only simulations, in which the smaller subhalos are more spherical than their counterparts in the field~\cite{Kuhlen:2007ku,Vera-Ciro:2014ita}.

\par The discussion above has made the assumption that the identified member stars in a dSph may be used as tracers of the local potential, and that they are not influenced by the potential of the Milky Way's dark matter halo. How justified is this assumption? The simplest metric to consider is whether or not the dynamical properties of a system change over a characteristic timescale. Characteristic timescales here are the orbital period of the dSph around the Milky Way halo, and the crossing time of stars within the dSph. From cosmological simulations, the former timescale is $\sim$ few Gyrs, and must include effects like mass loss, dynamical friction, and eccentricity as the dSph orbits around the Galaxy. If the crossing time is much shorter than this orbital period, a reasonably good approximation is to model the dSph as a stellar system in equilibrium in its own gravitational potential. Again, the best method to understand this question is via simulations; however given the present resolution of simulations the typical crossing time for a star is not as well measured as the orbital timescale. Satellites in simulations that are consistent with the brightest Milky Way dSphs may be tidally stripped in the outer parts of their halos, however when the core remains intact it is typically accurate to model the system as in equilibrium~\cite{Wang:2016qol}. 

\par With these basic theoretical principles in mind, for the remainder of this section we discuss the specific techniques that are used to model the dark matter distributions of the dSphs. We highlight in particular the recent work that addresses the various systematic uncertainties in the mass modeling.

\subsubsection{Distribution function} 

\par Kinematic information on a stellar system is fundamentally derived from the phase space distribution, $f$. For a system such as a dSph, in which the stars and the dark matter are distinct dynamical components, a separate phase space distribution describes each component. With this in mind define $f_\imath$ as the phase space distribution of a separate dynamical component, where $f_\imath = f_{DM}$, for the dark matter, and $f_\imath = f _\star$ for the stars. The density at ${\vec x}$  for a component is  
\begin{equation} 
\rho_\imath ({\vec x}) = \int f_\imath ({\vec x},{\vec v}) d^3 {\vec v}. 
\label{eq:df} 
\end{equation} 
This serves to normalize the distribution function so that the integral over velocities is the mass per unit volume. Assuming spherical symmetry the density is $\rho_\imath (r) = 4 \pi \int f_\imath (r,{\vec v}) d^3 {\vec v}$, and the velocity dispersion for a component is 
\begin{equation} 
\sigma_{\imath,\alpha}^2 (r) \equiv \overline{v_{\imath,\alpha}^2}= \frac{1}{\rho_\imath (r)} \int v_\alpha^2 f_\imath (r,\vec v) d^3 \vec v, 
\end{equation} 
where $\alpha = r, \theta, \phi$, the radial component is $\sigma_{\imath, r}^2$ and the tangential component is $\sigma_{\imath,t}^2 = \sigma_{\imath,\theta}^2 + \sigma_{\imath,\phi}^2$, where $\theta$ and $\phi$ are spherical coordinates. 

\par Define the energy of a particle as $E_\imath = v_\imath^2/2 + \Phi(r)$, where $v_\imath$ is the modulus of the velocity and $\Phi(r)$ is the total gravitational potential. The potential of the stars and the dark matter can be separately constructed numerically via the Poisson equation, $\nabla^2 \Phi_\imath = 4 \pi G \rho_\imath$. The total potential is then the sum of the two components. If the velocity dispersions are equal in the tangential and the radial directions, then the distribution function is isotropic, and the distribution function simplifies to a function of just the energy, $ f_\imath (E)$. With this assumption of isotropy, the distribution function is linked to the density profile via the Eddington formula, 
\begin{equation}
  f_\imath(E) = \frac{\sqrt{2}}{4\pi^2} \frac{d}{dE} \int_E^0
  \frac{d \rho_\imath}{d\Phi}\frac{d \Phi}{\sqrt{\Phi - E }}. 
  \label{eq:eddington}
\end{equation}
For an isotropic system, Equation~\ref{eq:eddington} is implies that given a model for the density profile of the stars and dark matter, and thus the total potential of the system, the phase space distribution is uniquely determined. 

%The density of a component is
%\begin{equation} 
%\rho_\alpha (r) = 2 \pi \int dv_r \int dv_t v_t f_\alpha(\epsilon,L) 
%\end{equation}

\par Recall that only the projected position and the line-of-sight velocity for stars are measured. This means that we must turn the distribution functions above into observed projected quantities. Defining $v_{||}$ as the line-of-sight velocity, for spherical and isotropic distribution functions this projection is
\begin{equation} 
\hat f(v_{||}, R)  = \int_R^{r_{||}} \frac{r dr}{\sqrt{r^2 - R^2}} \int_{(1/2) v_{||}^2 + \Phi(r)}^0 dE \, f(E)  
\label{eq:projf} 
\end{equation} 
%In terms of the distribution function, the observed quantity is the projected stellar distribution function. Including both energy and angular momentum dependence, this is given by, 
%\begin{equation} 
%\hat f(v_{los}; R)  = 2 \int_R^{r_{los}} \frac{r dr}{\sqrt{r^2 - R^2}} \int_0^{-(1/2) v_{||}^2 + \Phi(r)} d \epsilon \, g(\epsilon) \int_0^{2\pi} d\eta h(L) 
%\label{eq:projf} 
%\end{equation} 
where $2 \Phi(r_{||}) \equiv v_{||}^2$ and $R$ is the projected radius of the star. 
%Equation~\ref{eq:projf} represents and exact solution for the projected theoretical distribution function, for an input model of $g$, $h$, and the potential $\Psi$. 
From Equation~\ref{eq:projf} it is clear that even for the simplest case of spherical and isotropic models, several inputs are still required to model the projected distribution function, and from this determine the stellar distribution function. 

\par In order to gain insight into the form of the potentials that host the dSphs, we can appeal to simulations of dSph-like dark matter subhalos. In dark matter only simulations, the subhalos are well fit by Einasto density profiles~\cite{Springel:2008zz}. In combination with the dSph structural parameters and the stellar kinematics, the simulations can provide powerful insights into the nature of the dSph dark matter halos. Ref.~\cite{Strigari:2010un} identified dark matter subhalos in the dark matter-only Aquarius N-body simulations that are consistent with the dSph structural parameters and the stellar kinematics. Assuming isotropy, they determined the shape of the distribution function in Equation~\ref{eq:projf} using the potentials of the dark matter subhalos, and showed that the simulations are able to produce subhalos that match the dSphs. 

\par More general parameterizations for the distribution function of dSphs are possible, however it is not clear what the best motivated form of the distribution function is, especially for anisotropic models. Extending to account for anisotropic systems requires a model for the energy and angular momentum dependence of the distribution function.  For anisotropic and spherical systems, the distribution function depends on both the energy and angular momentum, $f_\imath(E,L)$, where the angular momentum modulus is $L = vr$. Assuming that the distribution functions are separable in energy and angular momentum, 
\begin{equation} 
f_\imath(E, L) \propto g_\imath(E) h_\imath(L). 
\label{eq:df_powerlaw}
\end{equation}
A mathematically-useful parameterization for the angular momentum distribution is $h(L) \propto L^{-2\beta}$. Assuming the $L^{-2\beta}$ parameterization, the anisotropy parameter is 
\begin{equation}
\beta =  1 - \frac{\sigma_t^2}{\sigma_r^2}. 
\label{eq:beta}
\end{equation}
Equation~\ref{eq:beta} may serve as a general definition of the velocity anisotropy; systems with more radial orbits have $\beta > 0$, and systems with more tangential orbits have $\beta < 0$. For anisotropic models, the additional functional dependance on angular momentum implies that a given density distribution may be produced from distribution functions with different functional dependence on $L$ and $E$. 

\subsubsection{Jeans modeling} 

\par Since the stellar and dark matter distribution functions may be complicated functions of energy and angular momentum, it is worthwhile to study modeling methods that do not rely on
assumptions for the phase space distributions of the stars or dark matter. Such a model may be constructed by taking moments of the distribution function. Assuming spherical symmetry and multiplying the collisionless Boltzmann equation by the radial velocity for the stars gives the Jeans equation, 
\begin{equation}
\frac{d (\sigma_{\star,r}^2 \rho_\star )}{dr} + \frac{2 \beta}{r} \rho_\star \sigma_{\star,r}^2 + \rho_\star \frac{d\Phi}{dr} = 0, 
\label{eq:jeans2}
\end{equation}
where $\sigma_{\star,r}^2$ is the radial velocity dispersion of the stars, and the stellar density profile is $\rho_\star(r)$. Binning the line-of-sight stellar velocities in projected radii $R$, and projecting the spherical Jeans equation along the line of sight gives 
\begin{equation} 
\sigma_{\star, ||}^2(R) = \frac{2}{I_\star(R)} \int_R^\infty \left[1-\beta(r)\frac{R^2}{r^2}\right] \frac{\rho_\star \sigma_{\star,r}^2 r}{\sqrt{r^2-R^2}} dr. 
\label{eq:jeans_los} 
\end{equation} 
Here the projected stellar density profile is $I_\star(R)$, which may be estimated from measurements of stellar photometry. The anisotropy function $\beta(r)$ is an input which requires an assumed functional form. 

\par Equation~\ref{eq:jeans2} and~\ref{eq:jeans_los} have been applied to dSphs for many years because of the simplicity of the input assumptions and the computational convenience. However, it is clear that there are several unknown parameters which must be constrained by the measured line-of-sight velocity dispersion and the stellar photometry. The degeneracies incurred implies that it is not possible to reconstruct the dark matter mass profile of a dSph from Equation~\ref{eq:jeans_los} alone. However, as mentioned above Equation~\ref{eq:jeans_los} does strongly constrain the integrated mass of the system within the half-light radius~\cite{Walker:2009zp,Wolf:2009tu} . This is reflected in Equation~\ref{eq:wolf}, which is found to be an excellent fit to a full Jeans analysis, nearly independent of the assumed anisotropy. 

\par The Jeans analysis may be further extended by taking higher order moments of the distribution function. In particular, at 4th order there are two Jeans equations that characterize the system, which are functions of the 4th order moments $\overline{v_{\star,r}^4}$, $\overline{v_{\star,t}^2 v_{\star,t}^2}$,  and $\overline{v_{\star,t}^4}$, 
\begin{eqnarray}
\frac{d( \rho_\star \overline{v_{\star,r}^4})}{dr} - \frac{3}{r} \rho_\star \overline{v_{\star,r}^2 v_{\star,t}^2} + \frac{2}{r}  \rho_\star \overline{v_{\star,r}^4} 
+ 3 \rho_\star \sigma_{\star,r}^2 \frac{d\Phi}{dr} &=& 0\\
\frac{d( \rho_\star \overline{v_{\star,r}^2 v_{\star,t}^2})}{dr} - \frac{1}{r}  \rho_\star \overline{v_{\star,t}^4} + \frac{4}{r} \rho_\star \overline{v_{\star,r}^2 v_{\star,t}^2}
+ \rho_\star \sigma_{\star,t}^2 \frac{d\Phi}{dr} &=& 0. 
\end{eqnarray}
Combining with equations for the 4th order moments completes the dynamical model for the system. However, unlike the case at 2nd order, solving this set of equations at 4th order requires an assumption for the stellar distribution function. Ref.~\cite{Lokas:2004sw} shows that for the assumption $h(L) \propto L^{-2\beta}$, the analysis further simplifies and the two Jeans equations may be combined into a single Jeans equation. Ref.~\cite{Richardson:2014mra} derived the moments assuming a Maxwell distribution for the components of the velocity. As discussed below in Section~\ref{sec:data}, the 4th order Jeans equations may be implemented in a maximum likelihood analysis similar to the 2nd order equation. 

\par With Jeans modeling it is also possible to understand the systematics behind the assumption of a spherically-symmetric potential.  Take $(R,\phi,z)$ as the standard cylindrical coordinates, assume that both the stars and dark matter are axisymmetric, and that the density distribution of the stellar system has the same orientation and symmetry as the dark matter. Define the anisotropy parameter as $\beta_z = 1 - \overline{v_z^2}/\overline{v_R^2}$, and take $\beta_z$ to be constant as a function of radius.  The solutions for the velocity dispersions are 
\begin{eqnarray}
 \overline{v_z^2} &=& \frac{1}{\rho_\star(R,z)} \int_z^\infty \rho_\star \frac{\partial \Phi}{\partial z} dz \\
\overline{v_\phi^2} &=& \frac{1}{1 - \beta_z} \left[  \overline{v_z^2} + \frac{R}{\rho_\star} \frac{\partial (\rho_\star  \overline{v_z^2})}{\partial R} \right]
 + R \frac{\partial \Phi}{\partial R}. 
\end{eqnarray} 
In order to compare to the measured dispersion profile, the $(R,\phi,z)$ coordinates must be converted to coordinates along the line-of-sight, which requires an assumption for the inclination of the object. With this theoretical model, the differences between the dark matter halo masses in spherical and specified non-spherical models have been discussed in Refs~\cite{Hayashi:2012si,Hayashi:2015yfa}. Implications for the J-factor are discussed below in Section~\ref{sec:results}. 

\par While both spherical and non-spherical Jeans methods are convenient, and as shown below can be easily implemented within a maximum likelihood analysis, there are some theoretical caveats to consider. Most importantly, given an assumed anisotropy profile, potential, and stellar density profile, there is no guarantee that this combination of functions results in a self-consistent stellar distribution function and thereby a physical solution. Further there may be systematics that are incurred in binning the data in projected radii, which is particularly important for dSphs with small data samples. Possible methods to mitigate these issues are discussed below in Section~\ref{sec:data}. 

\subsubsection{Orbit modeling}

\par  The orbit-based (or Schwarzschild modeling~\cite{1979ApJ...232..236S}) technique is probably the most general method that has been used to reconstruct the dSph potentials. The primary benefit of this method in comparison to the Jeans-based method is that orbit methods do not a priori require an assumption for the orbital anisotropy, or the form of the distribution function. Orbit-based methods have been recently applied to several dSphs with high quality data sets with a goal of determining their density profiles~\cite{Jardel:2011yh,Jardel:2012am,Breddels:2012cq}. Here we briefly discuss them in comparison to the other methods.  

\par Orbit-based models start by first creating a library of orbits that sample the available phase space in energy and angular momentum. For an assumed potential model, the orbits of the test particles are launched and integrated for a large number of crossing times. Each orbit is assigned a weight, and a set of weights are chosen to best match the photometry and kinematics.  Because the orbits are directly determined in this method, the velocity anisotropy and the phase space distribution for the best-fitting model is directly determined from the orbit distribution. This is in contrast to distribution function-based and Jeans-based methods,  which require assumptions for these functions. At this stage, the primary downside to orbit-based methods is that there is a significant computational cost to implement these models for the large modern data samples, so scanning over a large number of potential models isn't practical.  

\subsection{Photon flux from dark matter annihilation and J-factor} 
\par With the basic theory in place for determining dSph dark matter distributions, we are now in position to apply this theory to indirect dark matter detection. Starting very basic, the rate of dark matter annihilations is
\begin{equation}
d\Gamma = d \sigma \times | \vec{v}_1 - \vec{v}_2| \times \frac{\rho_{DM}}{m}, 
\label{eq:annihilationrate} 
\end{equation} 
where the velocities of the annihilating particles are $v_1$ and $v_2$, the differential cross section for annihilations is $d\sigma$, the dark matter density is $\rho_{DM}$, and the dark matter mass is $m$. Define the probability distribution of dark matter velocities as 
\begin{equation} 
P_{\vec x} = \frac{f_{DM}(\vec x, \vec v)}{\rho_{DM}(\vec x)}
\end{equation}
where the density is normalized as in Equation~\ref{eq:df}. The annihilation rate per unit volume is then 
\begin{equation}
\frac{d \Gamma}{dV} = \frac{\rho_{DM}}{m} \int {\rm d}^3 \vec v_1 {\rm d}^3 \vec v_2 P_{\vec x} (\vec v_1) P_{\vec x} (\vec v_2) (\sigma  | \vec{v}_1 - \vec{v}_2|) \times \frac{\rho_{DM}}{m}
\end{equation} 

\par Defining the center-of-mass velocity as $\vec v_{cm} = (\vec v_1 + \vec v_2)/2$ and the relative velocity as $\vec v_{rel} = \vec{v_1}-\vec{v_2}$, after integrating out the center-of-mass velocity, the above integral can be written as an integral over the relative velocity distribution, 
\begin{equation} 
\frac{d \Gamma}{dV} = \frac{\rho_{DM}^2}{m^2} \int {\rm d}^3 \vec v_{rel} P_{{\vec x}, rel} (\vec v_{rel}) (\sigma |\vec{v}_{rel}|), 
\end{equation} 
where the magnitude of the relative velocity is $v_{rel} = |{\vec v}_{rel}| = | {\vec v}_1 - {\vec v}_2 |$. This leads to the definition of the annihilation cross section averaged over the distribution function, 
\begin{equation} 
\langle \sigma v_{rel} \rangle =\int {\rm d}^3 \vec v_{rel} P_{{\vec x}, rel} (\vec v_{rel}) (\sigma |\vec{v}_{rel}|), 
\end{equation} 
which in general depends on spatial location ${\vec x}$. Along a direction specified by angles $\theta$ and $\phi$, and within a solid angle $\Delta \Omega$, the flux of photons is then 
\begin{equation} 
\frac{d \Phi}{d E} = \frac{1}{4 \pi m^2}  \frac{dN}{dE} \int d \Omega \int d \ell  \langle \sigma v_{rel} \rangle  \left[\rho_{DM}(r(\ell, \Omega))\right]^2 
\label{eq:flux}
\end{equation} 
where $dN/dE$ is the photon spectrum produced by an annihilation. 

\par It is often assumed that the annihilation cross section times the relative velocity, $\sigma v_{rel}$, is independent of the relative velocity. In this case, the standard J-factor within a solid angle $\Delta \Omega$ is defined as 
\begin{equation}
J(\Delta\Omega) = \int_{\Delta\Omega} d\Omega \int d\ell \left[\rho_{DM}(r(\ell, \Omega))\right]^2 \ , 
\label{eq:jstandard} 
\end{equation}
which depends only on the macroscopic distribution of dark matter in the dSph. 
%where the square of the dark matter density at $r$ is 
%\begin{equation}
%\left[ \rho_{DM}(r) \right]^2 =  \int  f_{DM}(r,\vec{v_1}) f_{DM}(r,\vec{v_2}) d^3 \vec{v_1} d^3 \vec{v_2}. 
%\end{equation}
With this assumption that $\sigma v_{rel}$ is independent of velocity, the flux does not depend directly on the phase space distribution of the dark matter, but rather can be derived from just a parameterization of the dark matter density profile, $\rho_{DM}(r)$. 

\par Most generally, $\sigma v_{rel}$ is velocity dependent, which implies that the flux in Equation~\ref{eq:flux} depends on $f_{DM}$. In this case the flux depends not only on the dark matter potential, but also on the orbital distribution of the particles. A simple assumption takes $f_{DM}$ to be a Maxwell-Boltzmann distribution, in which case $f_{DM}$ is independent of spatial position. The quantity $\langle \sigma v_{rel} \rangle$ may be pulled outside the integral in Equation~\ref{eq:flux}, the same definition for the J-factor applies as in Equation~\ref{eq:jstandard}, and 
 \begin{equation}
\langle \sigma v_{rel} \rangle = \int \sqrt{\frac{2}{\pi}} \frac{1}{v_p^3} v_{rel}^2 e^{-v_{rel}^2/2v_p^2} (\sigma v_{rel}) dv_{rel}
\end{equation} 
where $v_p$ is the most probable velocity. In addition to its theoretical simplicity, the Maxwell-Boltzmann assumption is convenient from a phenomenological perspective because $\langle \sigma v_{rel} \rangle$ can be bound from gamma-ray observations similar to the velocity-independent case.

\par If on the other hand $\langle \sigma v_{rel} \rangle$ depends on spatial position, then the definition of the J-factor depends on the assumed model for the phase space distribution. In this case, we must reconstruct the distribution function from either a set of theoretical assumptions, or by appealing to observations and implementing them with the context of the Eddington equation. As discussed below in Section~\ref{sec:results}, these models have been recently discussed in the interpretation of gamma-ray data~\cite{Ferrer:2013cla,Boddy:2017vpe}.  

\section{Data analysis} 
\label{sec:data} 
\par With the theoretical framework in place, we now discuss the statistical methods that are used for determining the best model for the dSph dark matter distributions. In this section we discuss in detail the different methods that are used, highlighting the strengths and weaknesses in each case. 

\par As discussed above the data come in the form of three phase space coordinates: the position on the sky and the line-of-sight velocity of a resolved star. The position of a star is typically measured to very high precision, so the uncertainties associated with it may be neglected. On the other hand, the uncertainties on the measured velocities must be accounted for in the statistical analysis. Model the observed velocity as the sum of the following contributions, 
\begin{equation}
v_\imath = u_\imath + e_\imath, 
\end{equation} 
where $u_\imath$ is the true velocity at the projected radius $R$ of the $\imath^{th}$ star that is predicted from the distribution function, and $e_\imath$ is the measurement uncertainty on the $\imath^{th}$ star. 
%The velocity $u_\imath$ comes from assumptions of the theoretical model. 
The true velocity is obtained from a probability distribution and is the sum of the following contributions, 
%The model velocity may not be directly determined from parameters that describe it, but rather it may be described by a probability distribution. In this case the model velocity is the sum of the following contributions,  
\begin{equation} 
u_\imath = \bar u + \sigma_\imath, 
\end{equation} 
where $\bar u$ is the mean velocity, and $\sigma_\imath$ is the intrinsic scatter at the projected radius of the $\imath^{th}$ star, $\sigma_\imath \equiv \sigma_{\star,||}(R_\imath)$. For the dSphs, most generally both the mean velocity $\bar u$ and the intrinsic scatter depend on the model parameters. For a spherically symmetric system, the model velocity is $\bar u = 0$. However in the data, even for a spherical system $\bar u$ is typically non-zero and reflects the systematic motion of the dSph relative to the heliocentric frame. 

\par For the measurement uncertainties $e_\imath$ are assumed to be Gaussian, 
\begin{equation} 
G(v_\imath,u_\imath,e_\imath) = \frac{1}{\sqrt{2 \pi e_\imath^2}} \exp \left[ - \frac{(v_\imath - u_\imath)^2}{2 e_\imath^2} \right]. 
\label{eq:gauss_measured}
\end{equation} 
The distribution for the model velocities $u_\imath$ may be more complicated, as it depends on the theoretical model for the distribution function. Starting off with a simple assumption that this distribution is Gaussian gives, 
\begin{equation} 
G(u_\imath,\bar u,\sigma_\imath) = \frac{1}{\sqrt{2 \pi \sigma_\imath^2}} \exp \left[ - \frac{(u_\imath - \bar u)^2}{2 \sigma_\imath^2} \right]. 
\label{eq:gauss_theory}
\end{equation} 
The probability for a measured velocity, given the model parameters, is then a convolution of Equation~\ref{eq:gauss_measured} and~\ref{eq:gauss_theory}
\begin{eqnarray} 
P(v_\imath | u_\imath, \sigma_\imath) &=& \int G(v_\imath,u_\imath,e_\imath) G(u_\imath,\bar u,\sigma_\imath) du_\imath \nonumber \\
&=& \frac{1}{\sqrt{2 \pi (\sigma_\imath^2 + e_\imath^2)}} \exp \left[ - \frac{(v_\imath - \bar u)^2}{2 (\sigma_\imath^2 + e_\imath^2)} \right]. 
\label{eq:pgauss}
\end{eqnarray} 
Here $u_\imath, \sigma_\imath$ implicitly depend on the model parameters. 

\par Now to be concrete consider a set of measured data, ${\cal V} = \{v_\imath \}$, and a set of model parameters ${\cal M}$. In the context of the discussion above, the model parameters are used to determine the projected velocity dispersion via the Jeans equation. These include parameters that describe the dark matter density profile, or the stellar velocity anisotropy. The probability for obtaining the data given the model parameters is $P({\cal V} |{\cal M})$. For a total of $n$ stellar velocity measurements, likelihood is then defined as, 
\begin{equation} 
{\cal L}({\cal M}) \equiv P({\cal V} |{\cal M}) = \prod_{\imath = 1}^n   \frac{1}{\sqrt{2 \pi (\sigma_\imath^2 + e_\imath^2)}} \exp \left[ - \frac{(v_\imath - \bar u)^2}{2 (\sigma_\imath^2 + e_\imath^2)} \right]. 
\label{eq:vlike} 
\end{equation} 
To obtain the posterior probability distribution function (pdf) for the parameters of interest, we can appeal to Bayes' theorem, 
\begin{equation} 
P({\cal M} | d) \propto P({\cal M}) {\cal L}({\cal M}),
\end{equation} 
where $P({\cal M})$ is the prior pdf on the model parameters. 

\par Equation~\ref{eq:vlike} is widely used to obtain best fitting model parameters in dSph kinematic analyses. It is particularly useful for dSphs with relatively small data sets, $< 100$ stars, and for cases in which the velocity dispersion of the system is low, $\sim 5$ km/s. For larger data sets the cost of computing the likelihood becomes significant. Further, if the model distribution deviates significantly from a Gaussian of Equation~\ref{eq:pgauss}, biases may be introduced in reconstructing the model parameters. Also note that it does not account for higher order contributions, such as binary star companion influence, that can affect the shape of the velocity distribution. 

\par Because of the possible bias induced with the assumption of gaussianity, variations on the above likelihood function have been implemented. For a large enough sample of stars, the velocities can be binned in projected radii, $\sigma_{||}^{obs}(R_\imath)$, and the data may be taken as the line-of-sight velocity dispersion profile as a function of $R$. In this case, the velocity dispersion is the estimator, and the likelihood may be taken as~\cite{Lokas:2003ks,Lokas:2004sw,Richardson:2014mra}
% and the probability for the velocity dispersion in the $\imath^{th}$ radial bin is, 
\begin{equation} 
{\cal L}({\cal M}) = \prod_\imath  \frac{1}{\sqrt{2 \pi \epsilon_\imath^2}} \exp \left[ - \frac{(\sigma^{obs}(R_\imath) - \sigma_{\star,||}(R_\imath))^2}{2 \epsilon_\imath^2} \right], 
\label{eq:psigma}
\end{equation} 
where the product is over the number of radial bins. Here $\epsilon_\imath$ is the standard error on the square of the velocity dispersion in the radial bin, which is derived from the sampling distribution. Here it is explicitly indicated that the velocity dispersion is determined at a radius $R_\imath$. 
%This is equivalent to the probability for the velocities in Equation~\ref{eq:pgauss}. 
Note that Equation~\ref{eq:psigma}, $\sigma_{\star,||}(R_\imath)$ is not necessarily the variance of a distribution that is assumed to be Gaussian. As above the posterior pdf for the model parameters such as the J-factor can be determined from Equation~\ref{eq:psigma}. Compared to the likelihood function constructed directly from the stellar velocities, the likelihood constructed from Equation~\ref{eq:psigma} is more computationally efficient for large data sets. 

%\par Equation~\ref{eq:pgauss} and ~\ref{eq:psigma} represent an approximation that the projected velocity distribution is gaussian at each projected radius $R$. This form of the probability distribution is convenient and therefore is often implemented in dSphs analyses. 

\par Though widely used, Equations~\ref{eq:vlike} and~\ref{eq:psigma} still ultimately represent and approximation to the likelihood function. The true likelihood function may be solved for by appealing to the model stellar phase space distributions, $f$, introduced in Section~\ref{sec:theory}. In this case the likelihood is modified as
\begin{equation} 
{\cal L} = \prod_\imath {\cal L}_\imath \propto \prod_\imath \int G(v_\imath,u_\imath,e_\imath)  \hat f(u_\imath, R_\imath)  du_\imath. 
\label{eq:calL} 
\end{equation} 
%with the normalization constant set by the condition $\int d v A f(v) = 1$. 
The model described by Equation~\ref{eq:calL} is theoretically more self-consistent than the models constructed from Gaussian assumptions for the likelihood function. However, the reason that this likelihood function is yet to be implemented for dSphs is that it is computationally expensive to scan a large region of parameter space. 

\par Note that all of the above analysis assumes that stars that are members of the dSph are perfectly identified, and there are no contaminating foreground stars. However, this is not typically the case, especially for foreground stars that have similar line-of-sight velocities as members stars. Understanding how to account for foreground stars requires a model for its probability distribution. A simple assumption to make about the foreground is that it is Gaussian, with a dispersion that is much wider than the assumed distribution for member stars~\cite{Ichikawa:2016nbi}. 

\par When obtaining the posterior pdf via a Bayesian method, the posterior depends strongly on the priors on model parameters, $P({\cal M})$~\cite{Martinez:2009jh}. For example, when characterizing the dark matter halo by the maximum circular velocity $V_{max}$ and the radius of maximum circular velocity $r_{max}$, the assumed prior on ($V_{max}, r_{max}$) strongly affects the resulting posterior. This is particularly relevant for dSphs with small samples ($< 100$) of stellar velocities. To try and alleviate this issue of priors for small data samples, Ref.~\cite{Martinez:2013els} introduced a Bayesian Hierarchical modeling method. This model takes as the data the measured value of the mass at the half-light radius, the measured half-light radius, and the luminosity of each dSph. The priors on ($V_{max}, r_{max}$) are not taken as an assumption but rather are determined from higher level parameters that describe the shape of the dark matter density profile (e.g. whether it is a Burkert, NFW, or Einasto model), and the parameters that describe the correlation between ($V_{max}, r_{max}$) that are taken from simulations. An additional part of this model is the relation between $V_{max}$ (or mass within a physical radius) and the dSph luminosity, which has been shown to be nearly flat over the entire range of dSph luminosities~\cite{Strigari:2008ib}. 

\par
\begin{figure}[h]
\includegraphics[width=15cm]{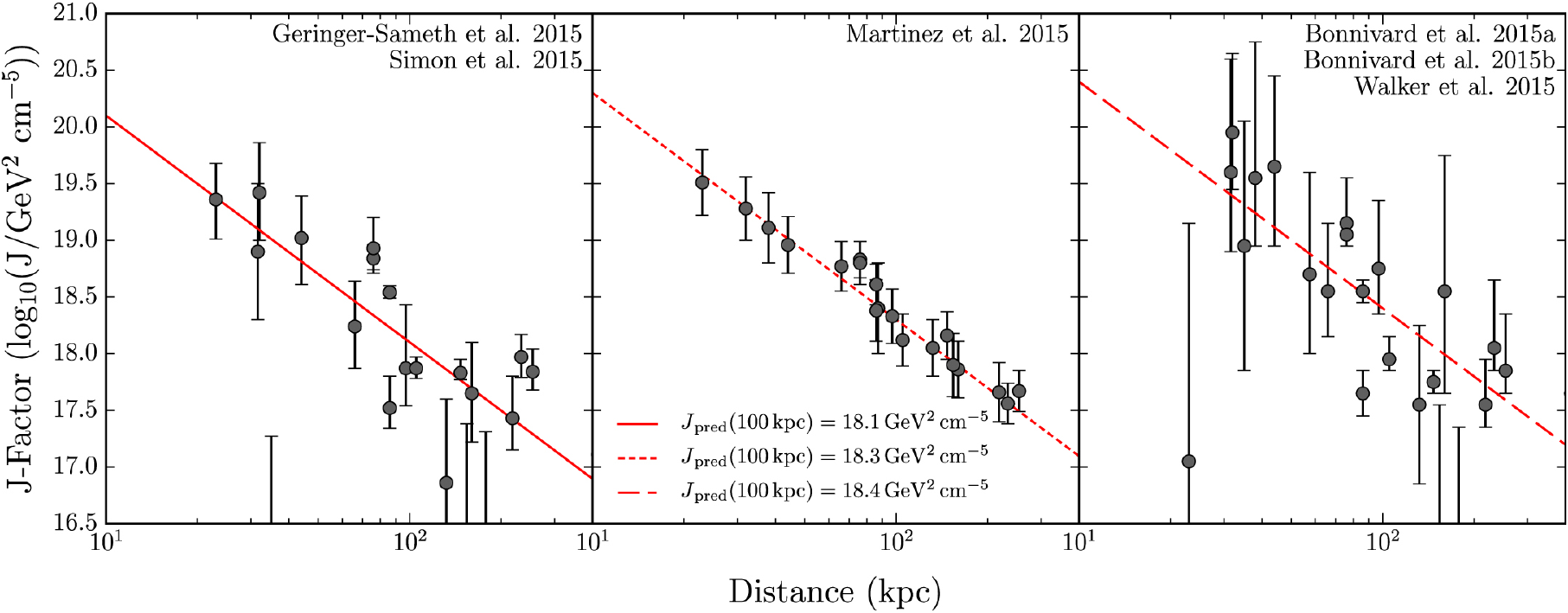}
\caption{J-factors as a function of distances, as calculated by the authors given in the upper right of each panel. The solid red lines are linear fits to the data. The left panel assumes flat, non-informative priors on the model parameters, the middle panel shows the results for the Bayesian hierarchical modeling method, and the right is similar to the left, except it assumes a more flexible model for the stellar velocity anisotropy and the stellar density distribution. Figure reproduced from Ref.~\cite{Fermi-LAT:2016uux}.} 
\label{fig:jfactor_comparison}
\end{figure}

 With the constraints on the parameters describing the dark matter mass distribution obtained from the likelihood methods outlined above, several authors have propagated this to determine the posterior pdf for the J-factors~\cite{Strigari:2007at,Essig:2010em,Charbonnier:2011ft,Geringer-Sameth:2014yza,Bonnivard:2015xpq}. Though these studies differ in specific assumptions for the dark matter profile, the anisotropy parameter, or in some cases the membership samples of stars used, the results are broadly in agreement, in particular for the dSphs with high quality data samples. The resulting J-factors as a function of distance are shown in Figure~\ref{fig:jfactor_comparison}. This figure compares calculations with different assumptions for Bayesian priors on model parameters. The left panel shows the case of flat, non-informative priors on model parameters, and the middle panel shows the results for the Bayesian Hierarchical modeling method. The right panel is similar to the left panel, but assumes a more flexible model for the stellar velocity anisotropy and the stellar density distribution. As this figure shows, the hierarchical modeling method of Ref.~\cite{Martinez:2013els} typically finds smaller uncertainties on the J-factor, mainly because of the priors assumed for ($V_{max}, r_{max}$). The right and left panels are in good agreement, with larger uncertainties on a typical J-factor because of the more flexible stellar profiles that are assumed. 

\par Complementing the above Bayesian methods, Ref.~\cite{Chiappo:2016xfs} recently introduced a frequentist-based method for parameter estimation using the likelihood functions. The frequentist approach has a couple of important motivations. First, it can avoid the systematic uncertainty introduced by the choice of priors in a Bayesian analysis. Second, it allows for a more self-consistent treatment of J-factor uncertainties in gamma-ray analyses, which mostly are frequentist. Ref.~\cite{Chiappo:2016xfs} finds that the J-factors and their uncertainties are generally in good agreement with the Bayesian-derived values, with the largest deviations restricted to the systems with the smallest kinematic data sets. The J-factors derived from these Bayesian and Frequentist methods are in good agreement with the analytic calculations of Ref.~\cite{Evans:2016xwx}. 

\par All of the above discussion assumes a spherically-symmetric model for the potential in combination with the likelihood functions. As discussed in Section~\ref{sec:theory}, even with this assumption there was a significant amount of theoretical freedom in the definition of the likelihood function, and the choice of which model parameters to vary, and over wide ranges. The situation only becomes much more complicated in the case of non-spherical models. 
%Recently, several authors have addressed the systematics issues that are incurred in estimating the J-factor with the assumption of spherical symmetry~\cite{Klop:2016lug,Hayashi:2016kcy,Sanders:2016eie}. 
%~\cite{Hayashi:2012si,Hayashi:2015yfa,Klop:2016lug,Hayashi:2016kcy,Sanders:2016eie}. 
A simple way to account for non-spherical models is to simply replace the spherical radius $r$ with the elliptical radius, 
\begin{equation} 
r = \sqrt{\frac{X^2}{a^2} + \frac{Y^2}{b^2} + \frac{Z^2}{c^2}} 
\end{equation} 
where $X, Y, Z$ are cartesian coordinates aligned along the principal axis of the dark matter halo. Depending on the projection axis along which the halo is viewed, the J-factor was shown to be biased, with the amount of bias depending on the sample size~\cite{Bonnivard:2014kza}. For these analyses, the Gaussian assumption for the likelihood function was employed, and so is subject to the same systematic uncertainties as the spherical models which make this assumption. 

\par Other methods for calculating J-factors with non-spherical models have been introduced. Ref.~\cite{Hayashi:2016kcy} consider the axisymmetric Jeans model as described in Section~\ref{sec:theory}, while Ref.~\cite{Sanders:2016eie} consider a flattened model for the stellar and dark matter distribution, and estimate the correction to the J-factor relative to spherical models for prolate and oblate dark matter haloes. Each of these models that utilize the axisymmetric Jeans equations requires model-dependent assumptions on the three-dimensional shape of the dark matter density distribution, and where necessary assume a Gaussian likelihood function for the stellar velocities.

\section{Indirect detection results} 
\label{sec:results} 
\par This section reviews the results from recent indirect dark matter searches that have targeted dSphs. The impact of the measured dark matter distribution and the J-factor is highlighted, in particular the systematics that these measurements induce in the interpretation of the limits on the dark matter annihilation cross section. Fermi-LAT results are highlighted first, followed by a discussion of results from Cherenkov observatories, and then a discussion of multi-wavelength studies. 

\subsection{Fermi-LAT results} 
\par The Fermi-LAT (Large Area Telescope) is an imaging, wide field-of-view pair conversion telescope that measures electron and position tracks that result from the pair conversion of incident high-energy gamma rays~\cite{Atwood:2009ez}. The LAT is sensitive to photons in the energy range $\sim 20 \, \textrm{MeV} -300 \, \textrm{GeV}$. In addition to the dSph results discussed in this section, the LAT has published a series of important indirect dark matter search results from a variety of astrophysical sources. Here we focus on the dSph results, classifying the LAT dSph analysis into two categories. First, the analysis for the dSphs with J-factors obtained directly from stellar kinematic measurements. Second, the analysis from probable dSphs which do not have measured stellar kinematics. For this second category, alternative methods must be used to estimate their dark matter distributions. 

\par As discussed above, dSphs have old stellar populations, and the dSphs that are used as indirect detection targets have no gas associated with them. In addition, the LAT dSph targets do not have any millisecond pulsars associated with them. Current best estimates of the gamma-ray flux from millisecond pulsars in dSphs find that for the brightest dSphs (Fornax and Sculptor) the probable flux is about an order of magnitude below the current LAT sensitivity limit~\cite{Winter:2016wmy}. For this reason, it is possible to neglect astrophysical sources of gamma-rays from the dSph itself when performing the dark matter analysis. 

\subsubsection{dSphs with stellar kinematics} 
\par Several bright dSphs have high-quality kinematic data sets, where here high quality means that the line-of-sight velocities of order 100 stars or more have been measured. These dSphs includes Fornax, Sculptor, Draco, Ursa Minor, Sextans, Carina, Leo I and Leo II. From the Jeans analysis discussed above, the stellar kinematic data  strongly constrains the integrated dark matter masses of these dSphs within their respective half-light radii of $\sim 100$ pc. 
%This result has been fleshed out by a number of authors over the past several years~\cite{Wolf:2009tu}. 
This result is important for the LAT analysis, because it implies that the constraint on the integrated mass within this region, and thus the J-factor~\cite{Walker:2011fs}, is very weakly dependent on whether the density distribution is modeled as a core or a cusp~\cite{Strigari:2007at}. 

\par The most promising dSphs for LAT analyses are those with the largest J-factor, which is sensitive to both the integral of the density squared and the distance. Amongst the dSphs with high quality data, Draco (80 kpc) and Ursa Minor (66 kpc) are the most promising. For these dSphs, the half-light radius corresponds to less than approximately one degree, which is approximately the angular resolution of the LAT for gamma-ray energies $\sim$ hundreds of MeV. More generally, independent of their mean measured J-factor, all of  the aforementioned dSphs with high quality data are important because the uncertainties on their J-factors are small, $\sim 0.1-0.2$ dex. 

\par In addition to the dSphs with high-quality data, several systems have smaller kinematic data sets. However, due to a combination of their proximity and dark matter distributions, 
as targets they may be just as promising as the dSphs with higher-quality data. Examples of these systems include Reticulum II, Segue I, Willman 1, Coma Berenices, and Ursa Major II. For these systems the dark matter distributions are estimated from smaller kinematic data sets of $\sim 10-60$ stars. There are two important issues that must be considered with these smaller data sets. First, even though in many cases the mean J-factor may be just as large as it is for systems such as Draco and Ursa Minor, the uncertainties on the J-factors are larger. Second, because of the smaller data samples, the astronomical nature of these objects may still be in question. For example, some of these objects may be more akin to globular clusters than galaxies, or it is even possible that these systems are being tidally stripped and not in dynamical equilibrium. 

\par Of the above dSphs with smaller data samples, the object with probably the least uncertainty in its astronomical classification is Reticulum II, which was discovered in the DES first year data. The velocity dispersion of this system is $\sim 3-4$ km/s~\cite{Simon:2015fdw,Walker:2015mla}, which implies a dark matter mass $\sim 500$ times larger than the stellar mass within the half light radius. Further, there appears to be no sign of tidal stripping of the stars.  Annihilation and decay factors for Reticulum II have been calculated by several groups, which are consistent for similar assumptions for theoretical priors~~\cite{Simon:2015fdw,Bonnivard:2015tta}. Reticulum II is also interesting because it has a small associated gamma-ray excess, though this is not statistically significant~\cite{Geringer-Sameth:2015lua,Hooper:2015ula,Zhao:2017pcz}.

\par Another interesting system that has recently been discovered in the DES and has been considered as a target is Triangulum II~\cite{Genina:2016kzg}. The first measurement indicated a velocity dispersion of $\sim 5$ km/s, with a factor of two rise to larger radii~\cite{2016ApJ...818...40M}. However, most recent measurements of 13 member stars constrain the velocity dispersion to $<$ 4 km/s, with no indication of a sharp rise at large radii. The initial results are likely explained by the contamination from binary stars~\cite{2017ApJ...838...83K}. 

\par Since the launch of the Fermi-LAT, several groups have searched for a gamma-ray signal from dSphs with stellar kinematic data~\cite{Ackermann:2011wa,Ackermann:2013yva,Geringer-Sameth:2014qqa,Ackermann:2015zua}.  No significant gamma-ray sources have been associated with the above dSphs. From these null observations, limits on the dark matter annihilation cross section, assuming that it is independent of relative velocity, can be obtained for a fixed value of the dark matter mass. These results, and nearly all subsequent results, assume that the dark matter annihilates entirely to a given set of SM particles. Because of the shape of the produced gamma-ray spectrum, the best limits come from annihilation into either $b \bar b$ or $\tau^+ \tau^-$. The results from an analysis of one year of data from the LAT showed that the upper bounds on the annihilation cross section approach the canonical thermal value for low mass dark matter, $\sim 10$ GeV~\cite{Abdo:2010ex}. 

\begin{figure}[h]
\begin{center}
\includegraphics[width=17cm]{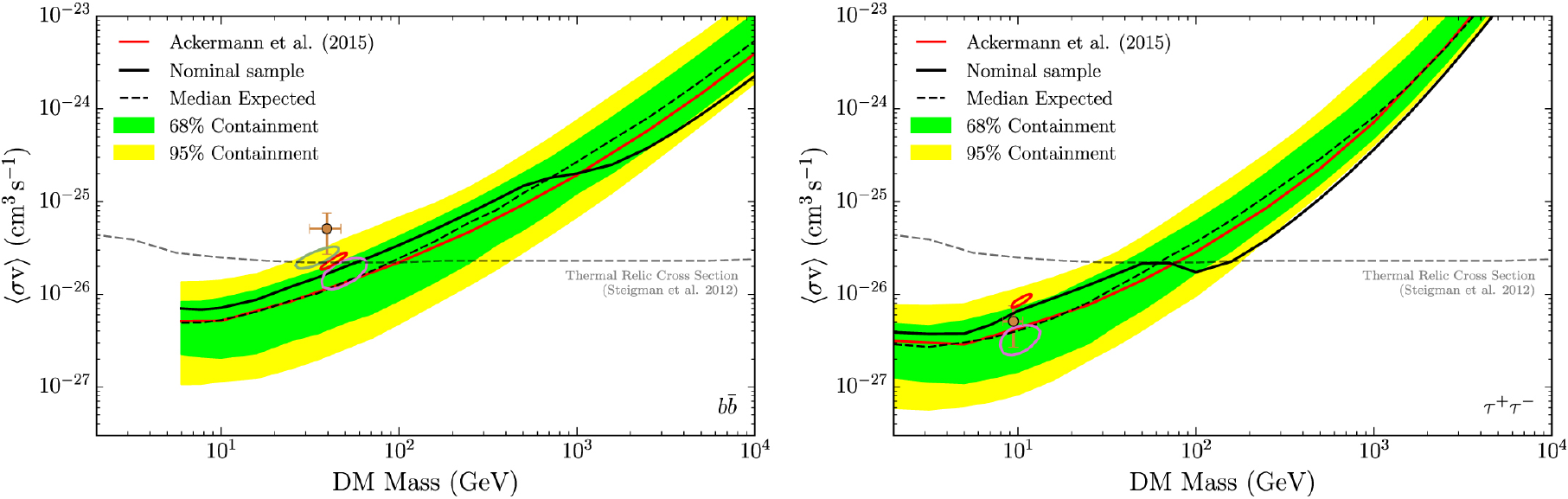}
\caption{Bounds on the annihilation cross section for $b \bar b$ and $\tau^+ \tau^-$ channels, compared to the thermal relic cross section (grey dashed). Bounds are shown from the sample with high quality stellar kinematics~\cite{Ackermann:2015zua} and from the entire satellite sample~\cite{Fermi-LAT:2016uux}. The dashed black curve shows shows the median expected limit from a blank sky gamma-ray analysis, and the green and yellow bands show the 68\% and 95\% quantiles. The ellipses and the point show uncertainty regions for a dark matter interpretation of the gamma-ray excess from the Galactic center. Figure reproduced from Ref.~\cite{Fermi-LAT:2016uux}.} 
\label{fig:gammaray_limits}
\end{center}
\end{figure}

\par The subsequent analyses of larger samples of LAT data combine the gamma-ray data from all of the dSphs with kinematic data. As is shown in Figure~\ref{fig:gammaray_limits}, this combined likelihood analysis provides stringent bounds on the cross section. In this combined analysis, Refs.~\cite{Ackermann:2011wa,Ackermann:2013yva,Ackermann:2015zua} constructed a full likelihood function in which the gamma-ray likelihood was multiplied by the likelihood for the J-factor. This full likelihood was then either profiled or marginalized to obtain the upper bound on the annihilation cross section~\cite{Ackermann:2015zua}. The posterior probability density for the J-factor was approximated as a log-normal distribution; this is a good approximation in particular for dSphs with high-quality data. 

\par The above procedure treats the J-factor as a statistical uncertainty in the likelihood analysis. There are additional systematic uncertainties that arise from the theoretical modeling of the stellar kinematics. Though as mentioned above the integrated dark matter mass within the half-light radius is well determined by the kinematic data, running through the entire analysis of all the dSphs when changing from an NFW model to a Burkert cored model can provide an $\sim 25\%$ change in the upper limit. Subsequently, Refs~\cite{Klop:2016lug,Hayashi:2016kcy,Sanders:2016eie} have quoted a similar systematic uncertainty when modeling non-spherical dark matter halos. 

\par In sum, analysis of dSphs with kinematic data has provided impressive limits on the annihilation cross section. The most recent Fermi-LAT combined results use six years of data to rule out thermal relic velocity-independent annihilation cross sections in the mass range $\sim 10-100$ GeV~\cite{Ackermann:2015zua}. As the analysis has progressed, the improvement in sensitivity has been a result of both the larger and cleaner gamma-ray data sample, and because the improvement in the analysis methods. For example, sensitivity to 100 GeV dark matter at the thermal relic cross section scales were achieved in part due to improvement in the spatial templates for the dSphs~\cite{Ackermann:2015zua}. Though systematic issues must still be addressed in this analysis pipeline, this is the most important result that the LAT has contributed to indirect dark matter searches. 

\subsubsection{dSphs with only photometry} 
\par In spite of the great progress that has been made measuring the kinematics and dark matter distributions of dSphs, many of the $\sim 40$ known Milky Way satellite galaxies do not have associated stellar kinematics. Most of the Milky Way satellites without stellar kinematics have been just recently discovered in the DES~\cite{Bechtol:2015cbp,Drlica-Wagner:2015ufc}. While many of these satellites have photometry characteristic of dark matter-dominated dSphs, it is only possible to confirm them as dSphs with kinematic follow up. 

\par Because the LAT is an all sky survey, all of these newly-discovered satellites can be followed up upon to determine if they are gamma-ray sources ~\cite{Drlica-Wagner:2015xua,Fermi-LAT:2016uux,Li:2015kag}. Though a few of the galaxies have small excesses above the background similar to that seen in Reticulum II, these excesses are not significant enough to associate a point source with them. Further, the satellites with the slight gamma-ray excesses do not correspond to the systems with the largest J-factors, providing further evidence that these small excess are not due to dark matter annihilation~\cite{Fermi-LAT:2016uux}. 

\par To derive upper limits on the the annihilation cross section from the null detection in these newly-discovered dSphs, a theoretical model must be developed for their dark matter distributions. For this modeling, we can start by considering the properties of the known dSphs with kinematic data. For dark matter-dominated dSphs, the kinematic data suggests that the central densities are very similar, even though they spread a wide range of nearly five orders of magnitude in luminosity~\cite{Strigari:2008ib}. From this result, combined with the distance to the dSph, the J-factor can be estimated, leading to an observed correlation between the J-factor and distance~\cite{Drlica-Wagner:2015xua} (Figure~\ref{fig:jfactor_comparison}). Assuming that the new satellites have J-factors that obey this relation, bounds on the cross section may be derived from the satellite population without known kinematics (Figure~\ref{fig:gammaray_limits}). While the above procedure does not substitute for obtaining high-quality stellar kinematics from dSphs, it does provide an estimate of how the LAT sensitivity can improve with the discovery of additional dark matter-dominated satellites. 

\subsubsection{Variations on Standard dSph analysis}
\par The above gamma-ray results from dSphs have made the simplest and most basic assumptions for both the dSphs and the nature of the dark matter annihilation into Standard Model particles. Several authors have extended upon this formalism to consider what the gamma-ray bounds imply for various other particle dark matter models. 

\par Probably most important, the limits obtained above have assumed that the annihilation cross section is independent of velocity. More generally, the velocity-dependent annihilation cross section can be parameterized as
\begin{equation} 
\sigma v_{rel} \simeq a + b v_{rel}^2, 
\label{eq:pwave}
\end{equation} 
where $a$ is the s-wave contribution, and $b$ is the p-wave contribution. Using the parameterization in Equation~\ref{eq:pwave}, Ref.~\cite{Zhao:2016xie} has used the combined likelihood method described above, and assumed a Maxwell-Boltzmann relative velocity distribution, to bound the annihilation cross section as a function of the ratio $a/b$, for different assumed annihilation channels. For a pure p-wave model ($a=0$), the upper bound is approximately three orders of magnitude weaker than the bound for pure s-wave model ($b=0$). 

\par While pure p-wave models suppress the annihilation cross section sensitivity because of the $v^2$ dependance, the sensitivity can be increased for Sommerfeld-enhanced models, which have annihilation cross sections that scale as $1/v^2$ or $1/v$. This scaling of the cross section may result from the exchange of a light mediator, or the formation of a bound state of the incoming dark matter particles. Again assuming a Maxwell-Boltzmann relative velocity distribution, the LAT data are very sensitive to the annihilation cross section, with upper limits of $\sim 10^{-30}$ cm$^3$ s$^{-1}$ for annihilation into $\tau^+ \tau^- \tau^+ \tau^-$~\cite{Zhao:2016xie}.

\par Because of the velocity dependance, the bounds on the annihilation cross section may be a sensitive function of the assumed dark matter phase distribution. The impact of the phase space distribution has been consider for Sommerfeld-enhanced models using both a Maxwell-Boltzmann distribution~\cite{Robertson:2009bh}, and self-consistent distributions derived from the constrained potentials~\cite{Ferrer:2013cla,Boddy:2017vpe}. The latter case is particularly intriguing, because the ordering of the J-factors from dSph to dSph may be different, depending on the constrained value of the potential and the precise shape of the velocity distribution~\cite{Boddy:2017vpe}.

\par The LAT dSph analysis is also sensitive to more complex annihilation channels. For example, if the dark matter annihilates to 4-body final states, such $b \bar b b \bar b$ or $\tau^+ \tau^- \tau^+ \tau^-$ or related combinations, the constraints may be altered because these models produce a larger fraction of photons at low energies~\cite{Dutta:2015ysa}. As discussed below this may have important implications for observed gamma-ray emission from the Galactic center. 

\par While the discussion above has focused mostly on the annihilation cross section, dSphs also provide bounds on the dark matter lifetime. In this case, the equivalent to the J-factor is simply an integral over the dark matter density along the line of sight. Using the combined likelihood method, for decay into such channels as $b \bar b$ or $\tau^+ \tau^-$, the lower bound on the lifetime for a $\sim 100$ GeV dark matter particle is $\sim 10^{26}$ s~\cite{Baring:2015sza}. 

\subsubsection{Related Fermi-LAT results} 
\par While the focus of this article is on dSphs, other galaxies in the Local Group have been studied by the LAT. Of particular interest is the Large Magellanic Cloud (LMC), which is the most luminous and most massive Milky Way satellite. The LMC is a dwarf irregular galaxy, and has a much more complicated structure and more recent star formation than the dSphs. The rotation curve of the LMC has been well-studied, through both gas motions and resolved stars. Combining theoretical models with the observed measurements of the dark mater distribution, Ref.~\cite{Caputo:2016ryl} place limits on the annihilation cross section that are comparable to those obtained from some dSphs. However, for the LMC there is a larger systematic uncertainty in the upper limit because of the larger uncertainty on its dark matter mass within the LAT field of view. 

\par M31 has also been a target of several LAT-relates studies~\cite{Fermi-LAT:2010kib,Ackermann:2017nya}. The most recent analysis indicates a nearly 10 sigma detection of gamma-rays from M31, and a nearly 4 sigma detection of an extended gamma-ray source. At this point, there is no clear consensus on the origin of these gamma-rays, as they do not correlate with regions of gas or star formation activity. These same authors have searched for gamma-ray emission from M33, a likely satellite galaxy of M31 with a stellar mass similar to that of the LMC. For M33, no gamma-ray signal was detected, and upper bounds were placed on the annihilation cross section given its estimated dark matter distribution.  

\subsection{Cherenkov observatories} 
\par The Fermi-LAT observations discussed above are complemented by observations at higher energies by Cherenkov telescopes. In these telescopes, Cherenkov light is detected from air showers that are produced in the atmosphere by gamma rays and cosmic rays. The typical effective area for these observatories is much larger than that of the LAT, $\sim 10^5$ m$^2$, and the angular resolution of approximately $0.1$ degrees over the appropriate energy regime is much better than that of the LAT. The lower energy thresholds are approximately hundreds of GeV, while the high energy thresholds of up to tens of TeV are much larger than that of the LAT. Cherenkov observatories are pointed instruments that are able to integrate for a long period of time on an individual dark matter source, though they are less sensitive to the diffuse radiation over large regions of the sky. In addition the sensitivity to a given dark matter cross section is typically a couple of orders of magnitude weaker than the sensitivity of the LAT. 

\par Because of their better angular resolution, Cherenkov observatories also differ from the LAT because they probe the central regions of the dark matter halos, within which there is a larger uncertainty on the integrated mass distribution. In this case, the J-factor depends more strongly on the assumed shape of the central dark matter profile, i.e. whether there is a core or a cusp. The systematics that are introduced because of this assumption have been quantified in Refs~\cite{Strigari:2007at,Essig:2009jx,Charbonnier:2011ft}. 

\par
\begin{figure}[h]
\begin{center}
\includegraphics[width=8cm]{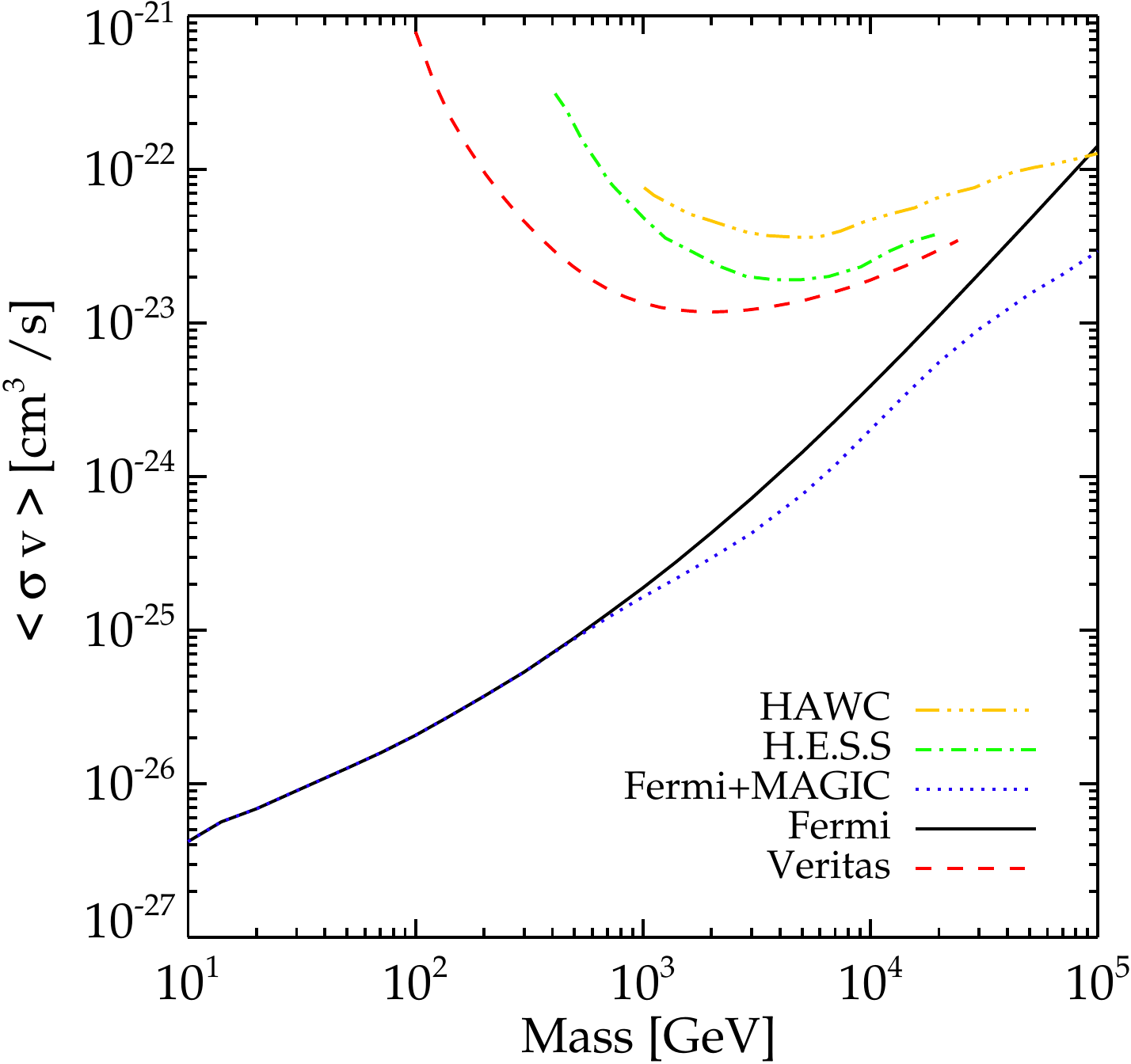}
\caption{Bounds on the annihilation cross section for $b \bar b$ from Cherenkov observatories. For comparison, the solid black curve is the Fermi-LAT only. The Fermi+ MAGIC results come from combining the Fermi bounds with MAGIC bounds from Segue 1. The H.E.S.S bounds from come observations of the Sagittarius dSph, and the Veritas bounds comes from a combined analysis of five dSphs. 
} 
\label{fig:cerenkov}
\end{center}
\end{figure}

 For the remainder of this subsection, we review the present bounds on dark matter in dSphs from the currently operating Cherenkov observatories. Figure~\ref{fig:cerenkov} summarizes the bounds for the example case of annihilation to $b \bar b$. 

\subsubsection{HAWC} 

\par The High Altitude Water Cherenkov (HAWC) telescope is located at Sierra Negra, Mexico. It is a wide field observatory sensitive to gamma rays in the energy range 500 GeV - 100 TeV, with an angular resolution of $\sim 0.5$ degrees. In this limit, it is accurate to approximate the dSphs as point sources.  HAWC presented results for 15 dSphs within its field of view~\cite{Albert:2017vtb}, using J-factors calculated from the code of Ref.~\cite{Bonnivard:2015pia} along with the best fitting parameters for the generalized NFW profile taken from Ref.~\cite{Geringer-Sameth:2014yza}. The best combined limits that HAWC obtains are for mass $\sim 1$ TeV  for annihilation to $\tau^+ \tau^-$, placing an upper bound on the annihilation cross section of $\sim 2 \times 10^{-24}$ cm$^3$ s$^{-1}$. 

\subsubsection{Veritas} 
\par The Very Energetic Radiation Imaging Telescope Array System (VERITAS) is an array of four imaging atmospheric Cherenkov telescopes. It is located at the Whipple Observatory in southern Arizona. VERITAS has excellent angular resolution, typically $< 0.1$ degree over the energy range 85 GeV - 30 TeV. For dark matter searches, VERITAS has targeted five dSphs: Bootes I, Draco, Segue 1, Ursa Minor, and Willman 1~\cite{Acciari:2010ab,Archambault:2017wyh}, using J-factors from Ref.~\cite{Geringer-Sameth:2014yza}. The best combined limits from VERITAS are for mass $\sim 300$ GeV, and for annihilation to $\tau^+ \tau^-$. At this mass the upper bound on the annihilation cross section is $\sim 3 \times 10^{-24}$ cm$^3$ s$^{-1}$. 

\subsubsection{Magic} 
\par The Florian Goebel Major Atmospheric Gamma-ray Imaging Cherenkov (MAGIC) is located in the Canary Island. MAGIC is an air Cherenkov telescope that is sensitive to gamma rays in the energy range $\sim 50$ GeV to 50 TeV. MAGIC has combined 158 hours of observations of Segue 1 with observations of 15 dSphs by the Fermi-LAT, and improved upon cross section limits obtained from each telescope alone by a factor of two for dark matter masses from $\sim 100$ GeV to 1 TeV~\cite{Aleksic:2011jx,Ahnen:2016qkx}.  For the $b \bar b$ channel, the upper bound on the annihilation cross section is $\sim 2 \times 10^{-25}$ cm$^3$ s$^{-1}$ at 1 TeV. 

\subsubsection{H.E.S.S.} 
The The High Energy Stereoscopic System (H.E.S.S.) is located in Namibia. Between 2006 and 2014, it obtained data on five dSphs: Sagittarius, Coma Berenices, Sculptor, Fornax, and Carina. Using J-factors determined Ref.~\cite{Martinez:2009jh}, H.E.S.S. obtained sensitivity to the annihilation cross section for all SM channels for dark matter masses ranging from approximately 400 GeV to 20 TeV~\cite{Abramowski:2014tra}. The best bounds come from the $\tau^- \tau^+$ channel for a dark matter mass of $\sim$ TeV; at this mass the upper bounds on the annihilation cross section is $\sim 3 \times 10^{-24}$ cm$^3$ s$^{-1}$. 

\subsection{Bounds at other wavelengths} 
\par While the focus above has been on gamma-ray results, dSphs have been studied at several other wavelengths at which a dark matter signal may arise. For example, to search for a signal associated with the injection of electron-positron pairs from dark matter annihilation, several authors have performed radio observations of nearby dSphs. Galaxies that have been targeted include Ursa Major II~\cite{Natarajan:2013dsa,Spekkens:2013ik}, Draco, Coma Berenices, and Willman 1~\cite{Spekkens:2013ik}, and Reticulum II~\cite{Regis:2017oet}. No radio signal has been associated with a known dSph, though new point sources have been identified that may be associated with gamma-ray point sources nearby to the respective dSphs. From these null observations, bounds can be placed on the annihilation cross section to charged particles, though the systematic uncertainty in the interpretation of these bounds is large due to the uncertainty in the magnetic fields associated with the dSphs. 

\section{Future} 
\label{sec:future} 
\par This article has reviewed the progress in measuring the dark matter distributions of dSphs, and the implications of these measurements for indirect dark matter detection. Over the past decade in particular, this field has evolved rapidly. The Fermi-LAT has achieved sensitivity to s-wave annihilation cross sections at the thermal relic scale, and since the mission is still operating, improvements on these bounds are anticipated. The LAT observations are complemented by Cherenkov telescopes with rapidly improving sensitivities at higher mass. In this section we discuss the path forward for studies of dark matter in dSphs, highlighting expected improvements in the near future. 

\subsection{Astronomical prospects} 

\subsubsection{Modeling dark matter distributions} 
\par There is substantial improvements still to be made in understanding the dark matter distributions of dSphs using line-of-sight velocity and stellar photometric data. As discussed above, these include implementation of self-consistent distribution function models using for example the likelihood function in Equation~\ref{eq:projf} for larger samples of line-of-sight velocities. Theoretical study will then be required to  extend this analysis to non-spherical models for the dark matter potential. Using the current analysis methods as implemented, probably in the near future the biggest gains will be increasing the kinematic samples of satellites discovered by the DES, up to plausible member samples of $\sim 50-100$. Samples of this size or even larger will be obtained by thirty-meter class telescopes, which will start collecting data within the next decade. 

\par An exciting method to obtain more information from present data sets, which has gained significant traction in the past several years, involves splitting up the kinematic samples into multiple stellar populations. Because stellar populations can be separated based on properties such as their metallicity, velocity, and spatial distribution, they may be viewed as distinct kinematic data sets. This is of great interest for the dSphs because the kinematic data can be reinterpreted as multiple data sets that probe the same underlying potential~\cite{Battaglia:2008jz,Walker:2011zu}. Several recents analyses have used multiple populations to study the dark matter potentials of dSphs, in particular Fornax and Sculptor, with the results obtained depending on the modeling method that is used~\cite{Walker:2011zu,Strigari:2014yea}. Both theoretical and observational improvements on this topic should happen in the next several years. 

\subsubsection{Proper motions}
\par Proper motions have long been discussed as a method to better understand dark matter profiles of dSphs~\cite{Wilkinson:2001ut,Strigari:2007vn}. In addition to the line-of-sight velocity, a measurement of the tangential velocity of stars in dSphs would provide a total of five out of the six phase space coordinates. In order to gain useful information on the dark matter distributions of dSphs, the required precision at the distance of a typical dSph is $\sim$ micro-arcsecond per year, which is several of orders of magnitude below the precision of modern measurements. This type of precision would have been possible with the SIM-Planetquest mission~\cite{Unwin:2007wj}, however unfortunately this was cancelled several years ago. A possible way forward with these types of measurements is using $\sim 10-20$ year baselines with the Hubble Space Telescope, though it has yet to be proven that the precision can be low as is required. Obtaining proper motions of bright stars in dSphs is also an important goal of thirty-meter class telescopes.

\subsection{Gamma-ray prospects} 
\subsubsection{Galactic center}  
\par Analyses of Fermi-LAT inner Galaxy data has revealed a diffuse gamma rays emission, which is nearly spherically-symmetric about the Galactic center~\cite{Goodenough:2009gk,Hooper:2010mq,Abazajian:2012pn,Daylan:2014rsa}. This emission is statistically significant, though its precise morphology and energy spectrum is still subject to systematic uncertainties~\cite{Cholis:2012am,TheFermi-LAT:2015kwa}. Pulsars~\cite{Abazajian:2010zy} or a population of point sources below the Fermi-LAT threshold~\cite{Bartels:2015aea,Lee:2015fea} have been fit to the data. Cosmic ray protons~\cite{Carlson:2014cwa} and inverse Compton emission from high energy electrons~\cite{Petrovic:2014uda,Cholis:2015dea,Gaggero:2015nsa} from  burst-like events may also explain this emission. 

\par Because of the uncertainties in the origin of these gamma rays from the inner Galaxy, a dark matter annihilation explanation of the Galactic center excess has generated considerable excitement. Dwarf spheroidals provide an ideal independent cross check on a dark matter interpretation of the emission from the inner Galaxy. Indeed at present the lack of excess gamma-ray signal from dSphs imposes constraints on annihilation cross-section, and also strongly constrains interpretations for a variety of different annihilation channels~\cite{Dutta:2015ysa}.  Future improvements in the gamma-ray limits from dSphs, and also possibly the discovery of new dSphs, are expected to go a way towards confirming or ruling out the dark matter interpretation of the gamma-rays from the inner Galaxy~\cite{Charles:2016pgz}. 

\subsubsection{Dark subhalos} 
\par  Simulations of Milky Way-like galaxies predict that $\sim 10-50\%$ of the mass is bound up in subhalos. The subhalos may be as small as the mass of the Earth, or even lower. Gamma rays from dark matter annihilation in these subhalos, the majority of which will not host stars, may also be detectable. 

\par Several ideas have been proposed to distinguish a dark matter subhalo from an unidentified astrophysical gamma-ray sources. A massive subhalo in the outer part of the Galactic halo or a much lower mass subhalo that is very nearby will have a spatial extension that is resolvable by the LAT. Also, if the source is a subhalo, its energy spectrum will deviate from a pure power law model. If the continuum gamma-ray spectrum is similar to a $b\bar b$ spectrum, subhalo sources may be distinguished from pure power law energy spectra~\cite{Ackermann:2012nb}. However, the energy spectra of pulsars typically has an exponential cut-off that mimics a $b \bar b$ spectrum, therefore based on the energy spectrum alone it is easy to confuse a potential subhalo source with a pulsar. Combining the extension and energy spectrum criteria with information on small variability, and cutting out sources near the Galactic plane where confusion is high, in the one and two year data no subhalo source could be uniquely identified~\cite{Zechlin:2011kk,Ackermann:2012nb,Zechlin:2012by}. For assumptions on the distribution of subhalos, this lack of detection corresponds to an upper limit on the annihilation cross section of approximately $10^{-24}$ cm$^3$ s$^{-1}$ at $100$ GeV. In addition to these upper limits, more recent analysis uses criteria outlined above to identify subhalos. Several candidate subhalos have been identified, but a conclusive confirmation still awaits~\cite{Bertoni:2015mla,Bertoni:2016hoh}.

\subsubsection{Cherenkov telescopes} 
\par The focus of future gamma-ray telescopes will be at energies above the energy sensitivity of Fermi-LAT. Experiments such as MAGIC, VERITAS, HESS, and HAWC will continue to take data and improve the current results. The Cherenkov Telescope Array (CTA)~\footnote{www.cta-observatory.org/}, which is expected to begin operation in the next few years, will extend dark matter limits to much higher masses, beyond the TeV scale, with better angular resolution than Fermi-LAT. CTA is expected to improve present limits from dSphs by at least an order of magnitude, and extend the Fermi-LAT studies of the Galactic center to a much higher energy regime~\cite{Doro:2012xx}. Ref.~\cite{Lefranc:2016dgx} have discussed the prospects for constraining the annihilation cross section with forthcoming CTA observations. 

\section*{Acknowledgements} 
I acknowledge support from NSF grant PHY-1522717 and DOE Grant de-sc0010813, and discussions with Leszek Roszkowski during the writing of this article. 

\bigskip 

\bibliographystyle{unsrt}
%\bibliography{main} 

\end{document}